\documentclass[aps,pra,twocolumn,showpacs,
amsmath,amssymb,floatfix]{revtex4}

\newcommand{\beq}{\begin{equation}}
\newcommand{\eeq}{\end{equation}}
\newcommand{\bea}{\begin{eqnarray}}
\newcommand{\eea}{\end{eqnarray}}

\begin{document}
\title{On the geometry of a class of $N$-qubit entanglement monotones}
\author{P\'eter L\'evay}
\affiliation{Department of Theoretical Physics, Institute of Physics, Budapest University of Technology and Economics, H-1111 Budapest Budafoki u 8.}
\date{\today}
\begin{abstract}
A family of $N$-qubit entanglement monotones  invariant under stochastic local
operations and classical communication (SLOCC) is defined.
This class of entanglement monotones includes the well-known examples of the concurrence, the three-tangle, and some of the four, five and $N$ qubit SLOCC invariants introduced recently.
The construction of these invariants is based  on bipartite partitions 
of the Hilbert space in the form ${\bf C}^{2^N}\simeq {\bf C}^{L}\otimes{\bf C}^{l}$ with $L=2^{N-n}\geq l=2^n$. Such partitions can be given a nice geometrical interpretation in terms of Grassmannians $Gr(L,l)$ of $l$-planes in ${\bf C}^L$ that can be realized as the zero locus of quadratic polinomials in the complex projective space of suitable dimension via the Pl\"ucker embedding.
The invariants are neatly expressed in terms of the Pl\"ucker coordinates of the Grassmannians.
\end{abstract}
\pacs{
03.67.-a, 03.65.Ud, 03.65.Ta, 02.40.-k}
\maketitle{}

\section{Introduction}
Since the advent of quantum information science \cite{Nielsen}  which regards entanglement as a resource it has become of fundamental
importance to characterize different classes of entanglement via the use of
suitable entanglement measures.
Though there are a number of very useful and spectacular results \cite{Popescu}, \cite{Wong}, \cite{Peres}, \cite{Horodecki}, \cite{Wootters} for quantifying the amount of entanglement present in pure and mixed states of multipartite systems, the subject is still at its infancy.  
For pure states for example we know that it is unlikely that the complete classification of $N$-qubit states will ever be found\cite{Luque} due to formidable computational difficulties.
Under such conditions it seems reasonable to try to find a characteristic subclass of $N$-qubit entanglement that can be described in a unified way. 
In this paper we attempt a modest step towards the identification of such a class which provides a way of understanding 
N-qubit entanglement in geometric terms.

The use of geometric ideas in understanding entanglement has already been used
in a number of papers\cite{Bengtsson,Brody,Mosseri,Bernevig,Miyake}.
In particular
it was observed\cite{Mosseri,Bernevig} that two and a special class of three-qubit entangled states can be described by certain maps that are entanglement sensitive. These maps enable a geometric description of entanglement in terms of fiber bundles. Fiber bundles are spaces which locally look like the product of two spaces the base and the fiber globally, however they can exhibit a nontrivial twisted structure. In this picture this twisting of the bundle accounts for some portion of quantum entanglement.
For two qubits these ideas were elaborated \cite{Lev1} using the correspondence between fibre bundles and the language of
gauge fields.
The essence of this approach  was to provide a description of entanglement by regarding the local unitary (LU) transformations corresponding to a {\it fixed} subsystem as gauge degrees of freedom.
In our recent paper \cite{Lev2} we have generalized this approach to describe the interesting geometry of three-qubit entanglement.
For this purpose we have taken into account the more general class of transformations corresponding to stochastic local operations and classical communication (SLOCC).
Using twistor methods we have shown that 
the relevant fibration in this case is a one over the Grassmannian $Gr(4,2)$
of two-planes in ${\bf C}^4$ with the gauge group being the SLOCC transformations of an arbitrarily chosen qubit i.e. $GL(2, {\bf C})$.
For every three-qubit state we have associated a {\it pair} of planes in ${\bf C}^4$, or equivalently a  {\it pair} of lines in the complex projective space ${\bf CP}^3$. 
In this picture entanglement can be described by the intersection properties of
a pair of lines in ${\bf CP}^3$.
Unlike the one in \cite{Bernevig} this method turned out to be capable of characterizing geometrically all the entanglement classes introduced in Ref. \cite{Dur}.
For example the two inequivalent classes of genuine three party entanglement
namely the GHZ and W classes correspond to the geometric situation of a pair of nonintersecting lines
or lines intersecting in a point respectively.

The aim of the present paper is to generalize these geometric ideas for multiqubit systems.  
We will see that for an interesting subclass of $N$ qubit entanglement such a generalization can indeed be done. 
The starting point of our investigations is a recent paper of Emary \cite{Emary}
introducing a class of entanglement monotones based on bipartite partitions
of multiqubit systems.
By reformulating and generalizing the results of Ref. \cite{Emary}
we are naturally led to a class of SLOCC entanglement monotones
giving back the well-known examples of the concurrence \cite{Wootters}, 3-tangle \cite{Kundu} and
some of the four \cite{Luque}, five \cite{Thibon} and $N$-qubit \cite{Wong} invariants introduced recently.
Moreover, these invariants can be rewritten in a nice and instructive form 
of geometric significance.
In fact these invariants are the natural ones associated to higher dimensional Grassmannians of $l$-planes that can be embedded in a complex projective space of suitable dimension. 
This observation leads us to the interesting possibility of understanding entanglement in  terms of the intersection properties of projective subspaces of a complex projective space of suitable dimension.
This approach being interesting and useful in its own right also shows a nice connection with twistor theory\cite{Lev2,Ward}.

\section{A bipartite class of entanglement monotones}

As a starting point we reformulate the results of Ref. \cite{Emary} in a geometric fashion convenient for our purposes.
Let us consider an arbitrary $N$-qubit pure normalized state $\vert \Psi\rangle\in {\bf C}^{2^N}$
\beq
\label{state}
\vert\Psi\rangle=\sum_{i_1,i_2,\dots i_N=0}^1C_{i_1i_2\dots i_N}\vert i_1i_2\dots i_N\rangle
\eeq
\noindent
where the states $\vert i_1i_2\dots i_N\rangle\equiv\vert i_1\rangle\otimes\vert i_2\rangle\dots\otimes\vert i_N\rangle$  correspond to the computational base of our $N$-qubit state.
Let us single out $n$ qubits such that $L\equiv 2^{N-n}\geq l\equiv 2^l$.
For convenience we chose these qubits to be the
{\it last} $n$ ones from the list $i_1i_2\dots i_N$, i.e. we have $i_1i_2\dots i_{N-n}i_{N-n+1}\dots i_N$.
Let us now construct the $L\times l$ matrix $Z_{\alpha a}, \alpha=0,1,\dots L-1, a=0,1,\dots l-1$ of $2^N=L\times l$ complex entries to 
be just $C_{i_1i_2\dots i_N}$ arranged according to this partition.
This means that the first $N-n$ terms of the binary string $i_1i_2\dots i_N$ written in decimal form are represented by $\alpha$ (rows) while the remaining $n$ terms in decimal form are represented by the letter $a$ (columns).
Since according to our assumption $N-n\geq n$ the matrix $Z_{\alpha a}$
is of rectangular shape with the number of rows is greater or equal than the number of columns. 

Let us assume now that the {\it columns} $Z_{\alpha 0},Z_{\alpha 1},\dots Z_{\alpha l-1}\equiv {\bf Z}_0,{\bf Z}_1,\dots {\bf Z}_{l-1}$ considered as {\it unnormalized } vectors in ${\bf C}^L$  are linearly independent. 
Then the matrix $(Z^{\dagger}Z)_{ab}$ (the reduced density matrix of the last $n$ qubits) is of maximal rank.
Hence the assumption of linear independence for {\it all} bipartite partitions is equivalent to the one that ${\vert\Psi\rangle}$ reinterpreted as the state of a {\it bipartite} system in ${\bf C}^{N-n}\otimes {\bf C}^n$ for all $N-n\geq n$  is totally entangled \cite{Ghirardi}.

Our unnormalized linearly independent vectors ${\bf Z}_0,{\bf Z}_1,\dots {\bf Z}_{l-1}$ span an $l$-plane in ${\bf C}^L$. The set of $l$-planes in ${\bf C}^L$ forms an $L-l\times l$ dimensional complex manifold the Grassmannian $Gr(L,l)$.
There are a number of ways to introduce complex coordinates for this manifold.
First the entries of the $L\times l$ matrix define the so-called homogeneous
or Stiefel coordinates. Their number is greater than the (complex) dimension
of the manifold. This redundancy in the homogeneous coordinates has its origin
in the fact that any linear combination of the vectors ${\bf Z}_a$, $a=0,1,\dots l-1$
spans the same $l$-plane. Equivalently, the transformation $Z\mapsto ZS$ where
$S\in GL(l,{\bf C})$ (the set of invertible $l\times l$ matrices with complex entries) can be regarded as a gauge degree of freedom. It merely amounts to
a redefinition of the vectors spanning the $l$-plane in question. 
It can be shown \cite{Gelfand} that $S(L, l)$ the set of complex $L\times l$ matrices $Z_{\alpha a}$
of full rank forms a fiber bundle over the Grassmannian $Gr(L,l)$ with gauge group, i.e. we have $Gr(L,l)=S(L,l)/GL(l,{\bf C})$.

Another way of defining homogeneous coordinates for $Gr(L,l)$
is to use the so-called Pl\"ucker coordinates.
By definition the Pl\"ucker coordinate $P_{{\alpha}_0{\alpha}_1\dots{\alpha}_{l-1}}$ of the $l$-plane defined by $Z$ is just the maximal minor of 
$Z_{\alpha a}$ formed by using the rows singled out by the $l$ fixed values ${\alpha}_0, {\alpha}_1,\dots {\alpha}_{l-1}$ .
It is obvious that if we make the transformation $Z\mapsto ZS$ with $S\in GL(l,{\bf C})$ the Pl\"ucker coordinates transform as
$P_{{\alpha}_0{\alpha}_1\dots{\alpha}_{l-1}}\mapsto Det(S)
P_{{\alpha}_0{\alpha}_1\dots{\alpha}_{l-1}}$.
The number of such coordinates is $\left({L\atop l}\right)$ which is greater than the dimension of the Grassmannian $Gr(L,l)$, this means that the Pl\"ucker coordinates are not independent. They are subject to special relations called the Pl\"ucker relations.

In order to illustrate these abstract concepts let us consider the example of a three-qubit system $N=3$. 
The state of the system can then be written in the form

\beq
\label{harom}
\vert\Psi\rangle=\sum_{{i_1,i_2,i_3}=0}^1C_{i_1i_2i_3}\vert i_1i_2i_3\rangle.
\eeq
\noindent
Let us chose $n=1$ corresponding to the {\it last} qubit, then we have $L=4$ ($\alpha =0,1,2,3)$ and $l=2$ ($a=0,1$) hence
\beq
\label{reszletes}
{\bf Z}_0\equiv
\begin {pmatrix} Z_{00}\\Z_{10}\\Z_{20}\\Z_{30}\end {pmatrix}=
\begin {pmatrix} C_{000}\\C_{010}\\C_{100}\\C_{110}\end {pmatrix}.
\eeq
\noindent
\beq
\label{reszletes2}
{\bf Z}_1\equiv
\begin {pmatrix} Z_{01}\\Z_{11}\\Z_{21}\\Z_{31}\end {pmatrix}=
\begin {pmatrix} C_{001}\\C_{011}\\C_{101}\\C_{111}\end {pmatrix}.
\eeq
\noindent
Now the Pl\"ucker coordinates are the maximal minors of the $4\times 2$
matrix $Z_{\alpha a}$ formed by the columns above.
Let us chose arbitrary two values ${\alpha}_0={\alpha}$ and ${\alpha}_1={\beta}$ then the Pl\"ucker coordinates are

\beq
\label{Pluckercoord}
P_{\alpha\beta}=Z_{\alpha 0}Z_{\beta 1}-Z_{\beta 0}Z_{\alpha 1}.
\eeq
\noindent
The number of such coordinates is $6$ which is greater then the complex dimension of $Gr(4,2)$ which is $4$.

Clearly under a $GL(2,{\bf C})$ transformation

\beq
\label{trans}
\begin {pmatrix} Z_{00}&Z_{01}\\Z_{10}&Z_{11}\\Z_{20}&Z_{21}\\Z_{30}&Z_{31}\end {pmatrix}
\mapsto
\begin {pmatrix} Z_{00}&Z_{01}\\Z_{10}&Z_{11}\\Z_{20}&Z_{21}\\Z_{30}&Z_{31}\end
{pmatrix}\begin {pmatrix} A&B\\C&D\end {pmatrix}
\eeq
\noindent
these coordinates transform as
\beq
\label{transpluck}
P_{\alpha\beta}\mapsto (AD-BC) P_{\alpha\beta}.
\eeq
\noindent
Hence although the number of Pl\"ucker coordinates is greater then the complex dimension of $Gr(4,2)$ 
we see from the equation above that these coordinates are defined merely projectively i.e. up to a nonzero complex number hence only their ratios count as coordinates. The number of ratios is $5$, moreover one can check that the quadratic Pl\"ucker  relation

\beq
\label{relation}
P_{01}P_{23}-P_{02}P_{13}+P_{03}P_{12}=0
\eeq
\noindent
holds which reduces the number of independent complex coordinates to $4$
the complex dimension of $Gr(4,2)$.

As we have seen the Pl\"ucker coordinates are defined up to a common scalar factor. Since these coordinates are defined merely projectively 
we should be able to embed
$Gr(L,l)$ into the complex projective space ${\bf CP}^D$ with $D=\left({L\atop l}\right)-1$. Such embedding really exists it is the Pl\"ucker embedding

\beq
\label{embedding}
Gr(L,l)\hookrightarrow {\bf CP}^{{\left({L\atop l}\right)}-1}={\bf P}\left(\bigwedge^l{\bf C}^L\right)
\eeq
\noindent
associating to the vectors ${\bf Z}_a$ , $a=0,\dots,l-1$ spanning the $l$-plane in question the {\it separable} $l$-vector ${\bf Z}_0\wedge{\bf Z}_1\wedge\dots\wedge {\bf Z}_{l-1}$ in the $l$-fold antisymmetric tensor product of ${\bf C}^L$ with itself.
For the three qubit case the Pl\"ucker embedding associates to the $2$-plane determined by the vectors ${\bf Z}_0$ and ${\bf Z}_1$ the separable {\it bivector}
${\bf Z}_0\wedge{\bf Z}_1$. Writing out this bivector as an antisymmetric matrix we get Eq. (\ref{Pluckercoord}).
Hence we can alternatively regard the Pl\"ucker coordinates as separable $l$-vectors
or as totally antisymmetric matrices with $l$ indices, satisfying additional constraints (Pl\"ucker relations).
In the language of $l$-vectors the transformation property of Pl\"ucker coordinates is

\beq
\label{transdet}
{\bf Z}_0\wedge\dots\wedge{\bf Z}_{l-1}\mapsto ({\rm Det S}){\bf Z}_0\wedge\dots\wedge{\bf Z}_{l-1},
\eeq
\noindent
where $S\in GL(l, {\bf C})$ is the usual $l\times l$ matrix acting on our $L\times l$ matrix $Z$. Clearly Eq. (\ref{transpluck}) is just a special case of (\ref{transdet}).

After illustrating our use of Pl\"ucker coordinates let us use them to express
the entanglement monotones of Ref. \cite{Emary} in a simpler form.
For this following \cite{Emary} let us introduce the operator $dx_{\alpha}$,
which assigns to vectors $\{{\bf Z}\}\equiv\{{\bf Z}_0,\dots{\bf Z}_{l-1}\}$ their  ${\alpha}$th component, i.e. $dx_{\alpha}({\bf Z}_a)=Z_{{\alpha}a}$, and combines them in the wedge product defined as

\beq
\label{wedge}
\bigwedge_{a=0}^{l-1}dx_{{\alpha}_a}(\{{\bf Z}\})={\rm Det}(dx_{{\alpha}_a}({\bf Z}_b))_{a,b=0,\dots l-1}.
\eeq
\noindent
Clearly this quantity is just the maximal minor of $Z$ labelled by the rows
${\alpha}_a, a=0,\dots l-1$, i.e the Pl\"ucker coordinate $P_{{\alpha}_1\dots{\alpha}_{l-1}}$.
In this notation the entanglement monotones 
\beq
\label{emary}
D^{(k_1,\dots k_n)}_n\equiv l^2\left(\sum_{{\alpha}_0<\dots<{\alpha}_{l-1}=0}^{L-1}
{\left|
\bigwedge_{a=0}^{l-1}dx_{{\alpha}_a}(\{\bf Z\})
\right|}^2\right)^{2/l}
\eeq
\noindent
of \cite{Emary} 
take the following instructive form

\beq
\label{em}
D^{(k_1,\dots k_n)}_n\equiv \frac{l^2}{l!}\left(\sum_{{\alpha}_0,\dots{\alpha}_{l-1}=0}^{L-1}
{\vert P_{{\alpha}_1\dots{\alpha}_{l-1}}\vert}^2\right)^{2/l}.
\eeq
\noindent
Notice that here we have introduced the general notation $(k_1,\dots k_n)$
of \cite{Emary}
to identify the location of the $n$ qubits. In our simplified case $(k_1,\dots k_n)=(N-n+1,\dots ,N)$ i.e. we have placed the $n$ qubits to the end of the $N$ qubit string. 
Clearly our considerations can be repeated for any partition with
$n$ qubit locations labelled as $(k_1,\dots,k_n)$ and a suitable adjustment for the definition of the Pl\"ucker coordinates for this case.
It should be obvious that for each such partition with fixed $L$ and $l$ we have a {\it different}
bundle of the form $S(L,l)/GL(l,{\bf C})$.
For a given $n$ we have $\left({N\atop n}\right)$ such entanglement monotones associated with these bundles, except for $n=N/2$ when we have half of this number.
The important property of the quantities
 $D^{(k_1,\dots k_n)}_n$ is that they are invariant under local unitary (LU)
 transformations of the qubits \cite{Emary}.
 Moreover, writing $P_{{\alpha}_1\dots{\alpha_{l-1}}}=({\bf Z}_0\wedge\dots\wedge{\bf Z}_{l-1})_{{\alpha}_1\dots{\alpha}_{l-1}}$ and using Eq. (\ref{transdet}) in (\ref{em}) we see that they are also invariant under the more general transformations
 of $U(l)$ acting on the $l$ qubit Hilbert subspace.  
Note that the quantities $D^{(k_1,\dots k_n)}_n$ are not necessarily independent.

\section{SLOCC entanglement monotones}

In order to motivate our generalization of the (\ref{em})
LU entanglement monotones
to SLOCC entanglement monotones we turn once again to the three-qubit case.
Let us single out the last qubit to be the one characterizing the partition.
Then we can write the antisymmetric matrix of Pl\"ucker coordinates in the form
${\bf P}={\bf Z}_0\wedge{\bf Z}_1$ i.e. as a separable bivector (see Eqs. 
(\ref{reszletes}),(\ref{reszletes2}) and (\ref{Pluckercoord})).
Then  we have $l=2$ and $L=4$ and the entanglement monotone $D_1^{(3)}$ can be written in the form

\beq
\label{tauc}
D_1^{(3)}=2\sum_{\alpha\beta=0}^3{\vert P_{{\alpha}{\beta}}\vert}^2
=4{\rm Det}\begin {pmatrix} \langle {\bf Z}_0\vert {\bf Z}_0\rangle&
\langle {\bf Z}_0\vert{\bf Z}_1\rangle\\  \langle{\bf Z}_1\vert {\bf Z}_0\rangle&\langle{\bf Z}_1\vert{\bf Z}_1\rangle\end {pmatrix},
\eeq
\noindent
where $\langle{\bf Z}_a\vert {\bf Z}_b\rangle\equiv \sum_{{\alpha}=0}^3\overline{Z}_{\alpha a}Z_{\alpha b}$, with the overbar denotes complex conjugation.
As it is well-known \cite{Kundu},\cite{Meyer},\cite{Brennen} 
$D_1^{(3)}={\tau}_{(12)3}=4{\rm Det}{\rho}_3=2(1-{\rm Tr}{\rho}_3^2)$ which is the linear entropy of the third qubit.
Repeating the same construction with the first and then the second qubit one gets the monotones $D_1^{(1)}$ and $D_1^{(2)}$ related to the linear entropies
of these qubits.
The quantity $Q_1=\frac{1}{3}(D_1^{(1)}+D_1^{(2)}+D_1^{(3)})$ is the permutation invariant used in \cite{Meyer} and \cite{Brennen}.

Let us now introduce a bilinear form $g:{\bf C}^4\times {\bf C}^4\to {\bf C}$
such that for two vectors ${\bf A},{\bf B}\in {\bf C}^4$
we have
\beq
\label{bili}
({\bf A}, {\bf B})\mapsto g({\bf A}, {\bf B})\equiv {\bf A}\cdot {\bf B}=
g_{\alpha\beta}A^{\alpha}B^{\beta}=A_{\alpha}B^{\alpha}
\eeq
\noindent
where
\beq
\label{gmatr}
g_{\alpha\beta}=g_{i_1i_2,j_1j_2}={\varepsilon}_{i_1j_1}{\varepsilon}_{i_2j_2},
\eeq
\noindent
or explicitly
\beq
\label{expl}
g=\begin {pmatrix} 0&0&0&1\\0&0&-1&0\\0&-1&0&0\\1&0&0&0\end {pmatrix}=
\begin {pmatrix} 0&1\\-1&0\end {pmatrix}\otimes
\begin {pmatrix} 0&1\\-1&0\end {pmatrix},
\eeq
\noindent
$\alpha,\beta =0,1,2,3$ and summation for repeated indices is understood.
Clearly $\overline{A\cdot B}=-\langle A\vert \tilde{B}\rangle$
where the right hand side is expressed via the spin flip operation of \cite{Wootters} i.e. $\vert\tilde{B}\rangle ={\sigma}_2\otimes {\sigma}_2\vert\overline{B}\rangle$. 

Let us now define the quantity similar to the one in (\ref{tauc})

\beq
\label{tangle}
{E}_1^{(3)}\equiv 2\vert P_{\alpha\beta}P^{\alpha\beta}\vert=
4\left|{\rm Det}\begin {pmatrix} {\bf Z}_0\cdot{\bf Z}_0&{\bf Z}_0\cdot{\bf Z}_1\\
{\bf Z}_1\cdot{\bf Z}_0&{\bf Z}_1\cdot{\bf Z}_1\end {pmatrix}\right|.
\eeq
\noindent
Notice the crucial changes we have made, namely we have taken the modulus of the sum, and the sum was understood
with respect to the metric (\ref{gmatr}).
Since $M{\varepsilon}M^t={\varepsilon}$ with $M\in SL(2, {\bf C})$ this sum with respect to $g$ is invariant under $SL(2, {\bf C})\times SL(2, {\bf C})$ i.e. 
of determinant one SLOCC transformations acting on the first and second qubits respectively.
Moreover, the (\ref{transpluck}) transformation property shows that the Pl\"ucker coordinates are invariant under the remaining $SL(2, {\bf C})$ transformation
of the third qubit.
Hence $E_1^{(3)}$ is an $SL(2, {\bf C})^{\otimes 3}$ invariant which can be shown \cite{Lev2} to be the three-tangle ${\tau}_{123}$ \cite{Kundu} which is also an entanglement monotone\cite{Dur}.
Moreover, it is easy to check that the invariants $E_1^{(1)}$ and $E_1^{(2)}$ defined similarly
are equal to $E_1^{(3)}$ reflecting the permutation invariance of the three-tangle.

Having gained some insight into the structure of three qubit invariants
now we turn to our generalization of SLOCC entanglement monotones.
(In the following by SLOCC transformations we mean the group $SL(2, {\bf C})^{\otimes N}$.)
The monotones we wish to propose are of the form

\beq
\label{ujmon}
E_n^{(k_1,\dots,k_n)}\equiv \frac{l^2}{l!}{\vert P_{{\alpha}_0\dots{\alpha}_{l-1}}P^{{\alpha}_0\dots{\alpha}_{l-1}}\vert}^{2/l},
\eeq
\noindent
where summation is now understood with respect to the $SL(2, {\bf C})^{\otimes (N-n)}$ invariant bilinear form with matrix

\beq
\label{altbili}
g_{\alpha\beta}={\varepsilon}_{i_0j_0}\otimes\dots\otimes{\varepsilon}_{i_{N-n-1}j_{N-n-1}}.
\eeq
\noindent
Hence the matrix of $g$ is just the $N-n$-fold tensor product of the fundamental $SL(2, {\bf C})$ invariant tensor ${\varepsilon}$.
An alternative formula using the $l$ linearly independent vectors
spanning the $l$-plane in question is
\beq
\label{monotone2}
{E}_n^{(\{k\})}\equiv
l^2{\left|{\rm Det}\begin{pmatrix}{\bf Z}_0\cdot{\bf Z}_0&\dots &{\bf Z}_0\cdot{\bf Z}_{l-1}\\ \hdotsfor[1.8]{3}\\ 
{\bf Z}_{l-1}\cdot{\bf Z}_0 &\dots &{\bf Z}_{l-1}\cdot{\bf Z}_{l-1} \end{pmatrix}
\right|}^{2/l}
\eeq
\noindent
In the following we adopt the convention of regarding the bilinear form $g$ to be fundamental, i.e. we consider the pair $({\bf C}^L, g)$ meaning that ${\bf C}^L$ is equipped with the extra structure defined by $g$.
Notice that for $N-n$ even the matrix $g$ is symmetric and for $N-n$ odd it is antisymmetric. For $N-n$ odd, $g$ defines a simplectic structure on ${\bf C}^L$.

The $SL(2, {\bf C})^{\otimes(N-n)}$ invariance of the quantities $E_n^{(\{k\})}$
($\{k\}\equiv(k_1,k_2,\dots,k_n)$ ) follows from the invariance of the bilinear form, and the $SL(2, {\bf C})^{\otimes n}$ invariance follows from the 
(\ref{transdet}) transformation formula of the Pl\"ucker coordinates
used for the subgroup $SL(2, {\bf C})^{\otimes n}\subset SL(l, {\bf C})$.
Hence the $ E_n^{({k})}$  are invariant under the full group of determinant one SLOCC transformations i.e. $SL(2, {\bf C})^{\otimes N}$.

The other important property of the quantities $E_n^{(\{k\})}$ is that they are entanglement monotones, meaning that on average they are non increasing under the action of any local protocol.
Now any local protocol can be decomposed into POVM (positive operator valued measures) acting on a single qubit. Since any POVM can be further 
be decomposed into a sequence of two-outcome POVMs, it is enough to demonstrate the non increasing property of the $E_n^{(\{k\})}$ under two outcome POVMs.
The proof that this property is indeed satisfied is simply a slightly modified rerun of the standard arguments that can be found in \cite{Dur,Emary,Moor}. 
The choice of the power $2/l$ in the (\ref{ujmon}) definition makes $E_n^{(\{k\})}$ to transform under local POVMs in the same way as the concurrence-squared
and the three-tangle do \cite{Emary}.

\section{Examples}

\subsection{Two and three qubits}

As our first example it is easy to show that in the case of two qubits ($N=2$, $n=1$, $L=l=2$) our entanglement monotones give back the usual definition of the concurrence squared. Indeed in this case $Z_{\alpha a}=C_{\alpha a}, (\alpha,a=0,1)$, hence we have a $2\times 2$ matrix $Z=C$ with linearly independent columns.
Two linearly independent vectors in ${\bf C}^2$ define the trivial Grassmannian $Gr(2,2)$ which is just a point.
For the monotone $E_1^{(2)}$ we have the formula

\beq
\label{conc}
E_1^{(2)}=4\vert{\rm Det} ({\bf Z}_a \cdot {\bf Z}_b)\vert=4\vert{\rm Det}(Z^tgZ)\vert=4\vert{\rm Det}C\vert^2,
\eeq
\noindent
which is just the concurrence squared. Clearly $E_1^{(1)}=E_1^{(1)}$.

For the three qubit case we have already shown that $E_1^{(1)}=E_1^{(2)}=E_1^{(3)}={\tau}_{123}$ with ${\tau}_{123}$ being the three-tangle.
Moreover, from Eq. (\ref{tangle}) we see that $E_1^{(3)}$ is just four times 
the magnitude of the discriminant of ${\rm Det}(xC_{i_1i_20}+yC_{i_1i_21})=0$, i.e. a binary form of degree two in the complex variables $x$ and $y$. 
According to the method of Schl\"afli\cite{Gelfand} this discriminant is just the hyperdeterminant $D(C)$ of $C_{i_1i_2i_3}$.

\subsection{Four qubits}

As our first nontrivial example let us consider an arbitrary four-qubit state

\beq
\label{four}
\vert{\Psi}\rangle=\sum_{i_1,i_2,i_3,i_4=0}^1C_{i_1i_2i_3i_4}\vert i_1i_2i_3i_4\rangle.
\eeq
\noindent
Let us first consider the partition $N-n=3, n=1$. In this case $L=8$ and $l=2$, hence for each four-qubit state totally entangled for this partition we have a $2$-plane in ${\bf C}^8$. Geometrically a four qubit state of this kind
determines a point in the Grassmannian $Gr(8,2)$, or equivalently 
a {\it line} in ${\bf CP}^7$.
Moreover the Grassmannian $Gr(8,2)$ as a manifold of complex dimension $12$ can be embedded in ${\bf CP}^{27}$ via the Pl\"ucker embedding.
In this case we have a $8\times 2$ matrix $Z_{\alpha a}$ with 
$\alpha =0,1\dots 7$ and $a=0,1$ consisting of the two columns

\beq
\label{reszl}
{\bf Z}_0\equiv
\begin {pmatrix} Z_{00}\\Z_{10}\\Z_{20}\\Z_{30}\\Z_{40}\\Z_{50}\\Z_{60}\\Z_{70}\end {pmatrix}=
\begin {pmatrix} C_{0000}\\C_{0010}\\C_{0100}\\C_{0110}\\C_{1000}\\C_{1010}\\C_{1100}\\C_{1110}\end {pmatrix}=
\begin {pmatrix} C_0\\C_2\\C_4\\C_6\\C_8\\C_{10}\\C_{12}\\C_{14}\end {pmatrix}
\eeq
\noindent
\beq
\label{reszl}
{\bf Z}_1\equiv
\begin {pmatrix} Z_{01}\\Z_{11}\\Z_{21}\\Z_{31}\\Z_{41}\\Z_{51}\\Z_{61}\\Z_{71}\end {pmatrix}=
\begin {pmatrix} C_{0001}\\C_{0011}\\C_{0101}\\C_{0111}\\C_{1001}\\C_{1011}\\C_{1101}\\C_{1111}\end {pmatrix}=
\begin {pmatrix} C_1\\C_3\\C_5\\C_7\\C_9\\C_{11}\\C_{13}\\C_{15}\end {pmatrix}
,
\eeq
\noindent
where for later use we also have written out explicitly the four qubit amplitudes using also decimal labeling.
The bilinear form on ${\bf C}^8$ is antisymmetric with the explicit form 

\beq
\label{g8}
g=\begin {pmatrix}0&0&0&0&0&0&0&1\\
                0&0&0&0&0&0&-1&0\\0&0&0&0&0&-1&0&0\\0&0&0&0&1&0&0&0\\ 
0&0&0&-1&0&0&0&0\\0&0&1&0&0&0&0&0\\0&1&0&0&0&0&0&0\\-1&0&0&0&0&0&0&0\end {pmatrix}.
\eeq
\noindent
Due to the antisymmetry of $g$ we have ${\bf Z}_0\cdot {\bf Z}_0={\bf Z}_1\cdot{\bf Z}_1=0$, hence the for the entanglement monotone $E_4^{(4)}$ we have the formula 

\beq
\label{e44}
E_1^{(4)}=4{\rm Det}{\left|\begin {pmatrix} 0&{\bf Z}_0\cdot{\bf Z}_1\\{\bf Z}_1\cdot{\bf Z}_0&0\end {pmatrix}\right|}=4{\vert{\bf Z}_0\cdot{\bf Z}_1\vert}^2.
\eeq
\noindent
We can also write this using the decimal labeling of the four qubit amplitudes
as

\begin{eqnarray}
\label{e442}
E_1^{(4)}=4\vert C_0C_{15}&-&C_2C_{13}-C_4C_{11}+C_6C_9-\nonumber\\ C_8C_7&+&C_{10}C_5+C_{12}C_3-C_{14}C_1\vert^2.
\end{eqnarray}
\noindent
Hence $E_1^{(4)}=4\vert H\vert^2$
where $H$ is the $SL(2,{\bf C})^{\otimes 4}$ invariant introduced in \cite{Luque}.
Calculating the invariants $E_1^{(1)}$, $E_1^{(2)}$ and $E_1^{(3)}$ by chosing the reduced qubits to be the first second and respectively the third a similar calculation shows that they are all equal to $E_1^{(4)}$ in accordance with the permutation invariance of $H$ \cite{Luque}. 
Later when we look at this invariant in a more general context we will give a simple proof of this fact.

Let us now calculate the invariant $E_2^{34}$. In this case
we have $N-n=2$ and $n=2$, hence $L=l=4$. 
In this case we have four vectors in ${\bf C}^4$ hence the Grassmannian
$Gr(4,4)$ being a point is again trivial. 
One then shows that 

\beq
\label{34matr}
Z_{\alpha a}=\begin {pmatrix} C_0&C_1&C_2&C_3\\C_4&C_5&C_6&C_7\\C_8&C_9&C_{10}&C_{11}\\C_{12}&C_{13}&C_{14}&C_{15}\end {pmatrix}.
\eeq
\noindent
Hence similar to the two-qubit case we have merely one Pl\"ucker coordinate
which is just the determinant of the matrix above, then we have

\beq
\label{L}
E_2^{(34)}=16\vert{\rm Det}({\bf Z}_a\cdot{\bf Z}_b)\vert^{1/2}=16\vert{\rm Det}Z\vert,
\eeq
\noindent
hence $E_2^{(34)}=16{\rm Det}\vert L\vert$ where $L$ is the $SL(2, {\bf C})^{\otimes 4}$ invariant introduced in \cite{Luque}. 
We can calculate two more invariants of this kind, namely $E_2^{(24)}$ and 
$E_2^{(14)}$ (the remaining ones are not independent).
A calculation shows that $E_2^{(24)}=16\vert M\vert$ and $E_2^{(14)}=16\vert N\vert$ in the notation of \cite{Luque}.
The $SL(2, {\bf C})^{\otimes 4}$ invariants $L$, $M$ and $N$ are still not independent due to the relation $L+M+N=0$.
Notice also that the same invariants arise from the ones of Emary, namely
$D_2^{(34)}$, $D_2^{(24)}$ and $D_2^{(14)}$ due to the fact that in this very special case the number of Pl\"ucker coordinates is merely one so the sums in (\ref{em}) and (\ref{ujmon}) contain merely one term (the sum of magnitudes in this case equals the magnitude of the sum).
Moreover, since the Hilbert series for the algebra of $SL(2, {\bf C})^{\otimes 4}$ invariants
is known\cite{Luque} it follows that the invariants $E_1^{(4)}$,
$E_2^{(34)}$ and $E_2^{(24)}$ are algebraically independent.
Moreover there are four invariants of degrees 2,4,4,6 generating freely the algebra of SLOCC invariants of a four qubit system. Our monotones already reproducing three of such fundamental invariants.

\subsection{Five qubits}

For a five qubit state
\beq
\label{five}
\vert{\Psi}\rangle=\sum_{i_1,i_2,i_3,i_4,i_5=0}^1C_{i_1i_2i_3i_4i_5}\vert i_1i_2i_3i_4i_5\rangle.
\eeq
\noindent
first we consider the partition $N=5$, $n=1$. In this case we have $L=16$ and $l=2$ so we have $2$-planes in ${\bf C}^{16}$. The set of such $2$-planes is the Grassmannian $Gr(16,2)$. Alternatively one can think of this space as the one
parametrizing the set of lines in ${\bf CP}^{15}$.
Now a five qubit state is characterized by the pair of vectors ${\bf Z}_0$ and ${\bf Z}_1$ forming the $16\times 2$ matrix $Z_{\alpha a}$  ($\alpha =0,1,\dots ,15, a=0,1$).
Now the invariant $E_1^{(5)}$ has the the same form as Eq. (\ref{tangle}) 
where now $g={\varepsilon}\otimes{\varepsilon}\otimes{\varepsilon}\otimes{\varepsilon}$.
Written out explicitly we see that the quantities ${\bf Z}_0\cdot{\bf Z}_0$ and ${\bf Z}_1\cdot{\bf Z}_1$ have the same structure as the one appearing in Eq. (\ref{e442}).
Indeed it is known that the invariant $H$ of degree two responsible for this structure defines a quadratic binary form in the variables $x$ and $y$.
The discriminant of this form defines an invariant of degree 4 \cite{Thibon}.
This discriminant is precisely of the (\ref{tangle}) form we are already familiar from the definition of the three-tangle via the use of Cayley's hyperdeterminant.
We can define four other invariants $E_1^{(1)}$, $E_1^{(2)}$, $E_1^{(3)}$ and $E_1^{(4)}$ similarly. One can show \cite{Thibon} that the invariants $E_1^{(j)}$
with $j=1,\dots 5$ are algebraically independent.

Let us now consider the partition $N=5$, $n=2$. In this case $Z_{\alpha a}$ is a $8\times 4$ matrix. Since $N-n=3$ is odd $g$ is antisymmetric, hence
${\bf Z}_a\cdot{\bf Z}_b=-{\bf Z}_b\cdot{\bf Z}_a$.
Hence the invariant $E_2^{(45)}$ has the form

\beq
\label{45inv}
E_2^{(45)}=16\vert{\rm Det}({\bf Z}_a\cdot {\bf Z}_b)\vert^{1/2}.
\eeq
\noindent
Since the determinant of an even dimensional antisymmetric matrix can always
be written as a square (the Pfaffian) we can write this as

\begin{eqnarray}
\label{452inv}
E_2^{(45)}=16\vert {\bf Z}_0\cdot{\bf Z}_1{\bf Z}_2\cdot {\bf Z}_3&-&\nonumber\\
{\bf Z}_0\cdot{\bf Z}_2{\bf Z}_1\cdot{\bf Z}_3&+&{\bf Z}_0\cdot{\bf Z}_3
{\bf Z}_1\cdot{\bf Z}_2\vert.
\end{eqnarray}
\noindent
Notice that there are $10$ entanglement monotones of this kind based on a partition of the form $5=3\oplus 2$. However, these invariants cannot be independent from the ones $E_1^{(j)}$ ($j=1,\dots, 5$) due to the results of \cite{Thibon} showing that the number of algebraically independent fourth order invariants is five.

\subsection{The $N$ qubit invariants of Wong and Christensen}

In a paper Wong and Christensen have introduced a potential entanglement measure calling it the $N$-tangle \cite{Wong}. 
In our notation they are just   
the invariants $E_1^{(N)}$ based on the partition $N=N-1\oplus 1$
corresponding to Grassmannians $Gr(2^{N-1},2)$ of $2$-planes in ${\bf C}^{2^{N-1}}$. 
In \cite{Wong} it was observed that for $N$ even these invariants can be written as a square of the pure state concurrence \cite{Wootters}.
This structure is indeed exhibited by our two and four-qubit invariants (\ref{conc}) and (\ref{e44}).
This result easily follows from the observation that the matrix $g$ of Eq. (\ref{altbili}) in this case is antisymmetric. 
Since the pure state concurrence is a permutation invariant we conclude that 
the invariants $E_1^{(N)}$ for $N$ even are also permutation invariants.
For the four qubit case we recover the well-known permutation invariance of $H$ of Ref. \cite{Luque}.

We also see that the invariants $E_n^{(\{k\})}$ arising from the partition $N=N-n\oplus n$ can always be written as a square of another invariant when $N-n$ is odd.
This again follows from the antisymmetry of $g$ and the fact that the determinant of an even dimensional antisymmetric matrix can be represented as a square of the Pfaffian. The simplest example of a Pfaffian is the combination (the Pl\"ucker relation) appearing in Eq. (\ref{452inv}).

\section{Conclusions}

In this paper we have introduced a class of $N$-qubit entanglement monotones
based on bipartite decompositions $N=N-n\oplus n$  of the Hilbert space ${\cal H}\simeq {\bf C}^{2^{N}}$.
This decomposition has naturally led us to the use of Grassmannians $Gr(L,l)$
of $l$-planes in ${\bf C}^L$ where $L=2^{N-n}\geq l=2^n$ as the natural structure characterizing the geometry of a subclass of $N$-qubit entanglement.
Our construction of such monotones was based on the paper of Emary \cite{Emary}. The new monotones unlike the ones in \cite{Emary} are SLOCC invariants, i.e.
invariant under stochastic operations and classical communication.
We have shown how the well-known invariants such as the concurrence, three-tangle, N-tangle and some of the four and five qubit invariants introduced recently
can be obtained as special cases.

There are a lot of interesting possibilities left to be explored.
The most important is of course to see what is the physical meaning of our monotones $E_n^{(\{k\})}$, for what kind of states we have
$E_n^{(\{k\})}=0$ etc. 
Moreover, an interesting development would be
the extension of the approach initiated in \cite{Lev2}
of characterizing different SLOCC classes of entanglement via studying
the intersection properties of $l-1$-planes in ${\bf CP}^{L-1}$.
A geometric approach of this kind would establish interesting links between the theory of entanglement and twistor theory \cite{Lev2, Ward}.
Such interesting questions will be addressed in a future publication.

\section{Acknowledgements}
Financial support from the Orsz\'agos Tudom\'anyos Kutat\'asi Alap
(grant numbers T047035, T047041, T038191) is
gratefully acknowledged.

\end{document}